\documentstyle[12pt]{article}

\advance \textheight by \topskip
\makeatletter
\@addtoreset{equation}{section}
  
\makeatother

\begin{document}

\title{\bf Hidden nonlinear supersymmetries in pure parabosonic systems}

\author{
Mikhail Plyushchay${}^{a,b}$\thanks{E-mail: mplyushc@lauca.usach.cl}\\
{\small ${}^{a}${\it Departamento de F\'{\i}sica, 
Universidad de Santiago de Chile,
Casilla 307, Santiago 2, Chile}}\\
{\small ${}^{b}${\it Institute for High Energy Physics, Protvino,
Russia}}}
\date{}

\maketitle
\vskip-1.0cm

\begin{abstract}
The existence of intimate relation between 
generalized statistics and supersymmetry is established
by observation of hidden supersymmetric structure
in pure parabosonic systems.
This structure is characterized generally by a nonlinear 
superalgebra. The nonlinear 
supersymmetry of parabosonic systems
may be realized, in turn, by modifying appropriately the 
usual supersymmetric quantum mechanics.
The relation of nonlinear parabosonic supersymmetry
to the Calogero-like models with exchange interaction
and to the spin chain models with inverse-square interaction 
is pointed out.
\vskip2mm
\noindent 
{\it PACS number(s): 11.30.Pb, 11.10.Lm, 03.65.Fd, 05.30.Pr}  
\end{abstract}

\newpage

\section{Introduction}
One of the fundamental and
seminal approaches in theoretical physics is 
a search for symmetries.
When new symmetry is revealed, this 
leads not only to better understanding of a system
but sometimes gives rise to establishing
unexpected  and important relations between different theories
and lines of research.

Generalized statistics first was introduced 
in the form of parastatistics as an exotic possibility 
extending the Bose and Fermi statistics
\cite{para01}-\cite{poly02} and for the long period of time
the interest to it was rather academic.
Nowadays it finds applications
in the physics of the quantum Hall effect
and (probably) it is relevant to high temperature
superconductivity \cite{any02}.
Supersymmetry, instead, unifies Bose and Fermi statistics 
\cite{susy01}-\cite{susyqm}
and its development lead to the construction of field
and string theories with exceptional properties \cite{susy03,susy04},
that transformed the same idea of supersymmetry
in one of the cornerstones of modern theoretical physics.
Though supersymmetry and generalized
statistics may be unified in the form of
parasupersymmetry \cite{rubspi}-\cite{parasu},
nevertheless, by the construction, the two concepts seem to be
independent.

In this paper we establish the existence of 
intimate relation between 
generalized statistics and supersymmetry
by observation of hidden supersymmetric structure
in pure parabosonic systems.
This structure is characterized generally by a nonlinear 
superalgebra. The nonlinear 
supersymmetry of parabosonic systems may be
realized, in turn, by modifying the 
usual supersymmetric quantum mechanics.
More specifically, we investigate hidden
supersymmetries of the single-mode parabosonic systems 
described by the Hamiltonians of the simplest quadratic
in parabosonic creation-annihilation operators
form and show that such simple systems possess a 
reach set of $N=1$ supersymmetries.
Depending on the order of a paraboson,
supersymmetry is realized either  
linearly in the unbroken or spontaneously broken
phase \cite{wit,susyqm}, or in the 
form of nonlinear symmetry \cite{n1}-\cite{n3}
generalizing for the supersymmetric case the known nonlinear
symmetries of the system of harmonic
bosonic oscillators with rational ratio of
frequencies \cite{gr}.
The basic tool with which we analyze 
the hidden supersymmetries of pure parabosonic
systems is the so called $R$-deformed Heisenberg algebra
(RDHA)\cite{vas}-\cite{def}, introduced inexplicitly by Wigner under
generalization of the bosonic oscillator \cite{para} 
that lead subsequently to the concept of parafields
and parastatistics \cite{para01,para,poly02},\cite{poly01}-\cite{paras}.
Due to the RDHA structure underlying parabosons,
as we shall see,
the systems under investigation
turn out to be related to inverse-square interacting particle
systems \cite{poly02,in1,poly},
which are important models of many-body systems
because of their exact solvability and intimate
connection to spin chain systems \cite{spch1}
and 2-dimensional Yang-Mills theories \cite{YM1}.
Recently, this special deformation of Heisenberg algebra 
has found applications for the description
of 3D \cite{mp00}-\cite{kuz} and 2D \cite{2d} anyons,
3D supersymmetry \cite{3ds}
and higher-spin gauge interactions \cite{v1}. 
It lies also in the basis 
of the construction of the minimally bosonized 
version of supersymmetric
quantum mechanics \cite{mp01,mp02,gpz}. 

One of the basic ingredients of supersymmetric quantum mechanics 
\cite{wit,susyqm}
is the $Z_2$-grading operator $\Gamma$, $\Gamma^2=1$, which
classifies all the operators into 
even (bosonic, $B$) and odd (fermionic, $F$)
subsets according to the relations  
$[\Gamma,B]=\{\Gamma,F\}=0$.
The main idea of the minimally bosonized supersymmetric
quantum mechanics 
is to take the reflection (parity) operator $R$, 
$
R\Psi(x)=\Psi(-x),
$
as a grading operator.
With this, the operators 
\begin{equation}
Q_1=iRQ_2,\quad
Q_2=-\frac{i}{\sqrt{2}}\left(\frac{d}{dx}+W(x)R\right),
\label{qq12}
\end{equation}
\begin{equation}
H=\frac{1}{2}\left(-\frac{d^2}{dx^2}+W^2(x)-W'(x)R\right),
\label{hams}
\end{equation}
containing an odd function $W(x)=-W(-x)$,
are the odd supercharges, $\{R,Q_i\}=0$, $i=1,2$,
and the even Hamiltonian,
$[R,H]=0$,
satisfying the linear superalgebra of Witten's
supersymmetric quantum mechanics:
\begin{equation}
\{Q^+,Q^-\}=H,\quad
Q^{\pm 2}=0,\quad
[H,Q^\pm]=0,
\label{susyal}
\end{equation}
where $Q^\pm=\frac{1}{2}(Q_1\mp iQ_2)$.

With the choice of the superpotential $W(x)=-\frac{\nu}{2x}$,
$\nu\in {\bf R}$,
the operator $\sqrt{2}Q_2$ coincides with the Yang-Dunkl
operator \cite{yd} 
\begin{equation}
D_\nu=-i\left(\frac{d}{dx}-\frac{\nu}{2x}R \right),
\label{yd}
\end{equation}
related to the Calogero model \cite{poly,macf},
where $R$ plays the role of the exchange operator.
With the extended differential operator $D_\nu$,
one can construct the analogs of bosonic
creation-annihilation operators,
\begin{equation}
a^\pm=\frac{1}{\sqrt{2}}(x\mp iD_\nu).
\label{aadef}
\end{equation}
These operators together with the reflection operator
$R$ form the $R$-deformed Heisenberg algebra,
\begin{equation}
[a^-,a^+]=1+\nu R,\quad 
\{R,a^\pm\}=0,\quad
R^2=1,
\label{dha}
\end{equation} 
which possesses unitary infinite-dimensional 
representations for $\nu>-1$  and 
at the integer values of the deformation parameter,
$\nu=p-1$, $p=1,2,\ldots$, 
is directly related to parabosons of order
$p$ \cite{para,param,def}. On the other hand, at $\nu=-(2p+1)$  
the algebra (\ref{dha}) has finite-dimensional representations
corresponding to the deformed parafermions of
order $2p$ (see Appendix).

Earlier it was observed \cite{mp02} that 
the systems given by the Hamiltonians
\begin{equation}
H_\epsilon=\frac{1}{2}\{a^+,a^-\}-\frac{1}{2}\epsilon
R[a^-,a^+],\quad 
\epsilon=+,-,
\label{hamdef}
\end{equation}
are supersymmetric: they are described by Eqs. (\ref{qq12}),
(\ref{hams}) with $W=\epsilon x -\frac{\nu}{2x}$.
These systems possess the following property:
in the case $\epsilon=+$ the spectrum of the Hamiltonian
(\ref{hamdef}) does not depend on the value of the 
deformation parameter $\nu>-1$, i.e. Eq. (\ref{hamdef})
gives the isospectral family of supersymmetric systems
being in the phase of exact supersymmetry ($E_0=0$).
On the other hand, at $\epsilon=-$
the Hamiltonian (\ref{hamdef}) describes the set of systems 
in the phase of spontaneously broken supersymmetry
with the scale of supersymmetry breaking $E_0=1+\nu>0$
governed by the value of the deformation parameter $\nu>-1$.

Here we first show that at $\nu=1$
the unbroken and spontaneously broken 
supersymmetries of the system
(\ref{hamdef}) are the linear supersymmetries of
the parabosonic systems  described by the Hamiltonians of the simplest form
$H_n=a^+a^-$ and $H_a=a^-a^+$.
Then we shall find that at $\nu=2k+1$, 
$k=1,2,...,$ the same quadratic Hamiltonians 
supply us with a supersymmetry
characterized by nonlinear superalgebra. The
peculiarity of linear and nonlinear supersymmetries
of pure parabosonic systems, as we shall see, is
encoded in the nature of their
supercharges: being realized in terms of parabosonic
creation-annihilation operators $a^\pm$,
they are essentially nonlinear.  We shall 
observe that at the values of deformation parameter $\nu=2k$
corresponding to the parabosons of order $p=2k+1$, the
spectra of the Hamiltonians $H_n$ and $H_a$ have no supersymmetry
but possess the interesting structure: they reveal a finite
number of `holes' in comparison with bosonic spectra ($\nu=0$).
This `hole number' is correlated with the parastatistics' order.  
It will be shown that with 
the  single-mode supersymmetric parabosonic systems
one can associate the systems of  
two identical fermions, which are the simplest two-particle
spin chain models with inverse-square interaction. 
We shall also discuss
the realization of parabosonic supersymmetries by
modifying the classical analog of usual supersymmetric
quantum mechanics.

The paper is organized as follows.  Section {\bf 2} is devoted to
investigation of the supersymmetries of the parabosonic systems given by 
the normal ordered Hamiltonian $H_n=a^+a^-$. 
Section {\bf 3} deals with the systems
described by the Hamiltonian $H_a=a^-a^+$.  Here, in particular,
we find the exotic parabosonic system with the spectrum of usual 
superoscillator \cite{nic}
but characterized by the nonlinear superalgebra.  
Section {\bf 4}
concerns the realization of the parabosonic
supersymmetries in the system of two identical fermions.
In Section {\bf 5} we show how the nonlinear
supersymmetry of pure parabosonic systems 
may be reproduced by appropriate modification
of the classical analog of the usual supersymmetric 
quantum mechanics.
In concluding Section {\bf 6} we
discuss the revealed supersymmetries
from the view-point of generalized deformed oscillator
\cite{das}, Calogero-like models with exchange
interaction \cite{poly02,poly,poly03},
reduced parasupersymmetry,
and nonlinear finite $W$-symmetries \cite{n3}
and give some list of open problems for further 
investigation.
Appendix  contains the necessary information
on irreducible representations of RDHA and on the 
relationship of RDHA to parabosons and parafermions.  

\section{The systems with Hamiltonian $H_n=a^+a^-$}

Let us choose the superpotential in Eqs. (\ref{qq12}),
(\ref{hams}) in the form
\begin{equation}
W=x-\frac{1}{2x}.
\label{nu1}
\end{equation}
In this case the Hamiltonian (\ref{hams}) 
can be represented as the normal ordered 
product
\begin{equation}
H_n=a^+a^-
\label{norm}
\end{equation}
of the operators (\ref{aadef}) corresponding to the particular
value of the deformation parameter: $\nu=1$.
This form of the Hamiltonian can also  be obtained
from Eq. (\ref{hamdef}) by putting in it $\epsilon=+$ 
and $\nu=1$.
The supersymmetric Hamiltonian (\ref{norm})
may be written equivalently as 
$
H_n=N+\Pi_-,
$
where $N$ is a number operator,
$[N,a^\pm]=\pm a^\pm$, 
and $\Pi_-=\frac{1}{2}(1-R)$ is a projector
(see Appendix).
{}From here the spectrum of the normal ordered Hamiltonian
(\ref{norm}) with $a^\pm$ corresponding to the case $\nu=1$ 
of algebra (\ref{dha})
can be immediately obtained:
$H_n\vert l\rangle =E_l\vert l\rangle$,
$E_0=0$, $E_{2l-1}=E_{2l}=2l$, $l=1,2,\ldots$.
So, the bosonized supersymmetric system 
characterized by superpotential
(\ref{nu1}) is the single-mode 
parabosonic (of order $p=2$) system 
with unbroken  supersymmetry.

Let us take the normal form for the Hamiltonian
(\ref{norm}) in general case of unitary representations
of RDHA ($\nu>-1$) and investigate its spectrum.
In terms of the number operator 
we have 
\[
H_n=N+\nu\Pi_-.
\]
The even states $\vert 2l\rangle$
are stable eigenstates of $H_n$:
their eigenvalues do not depend on 
the deformation parameter,
\[
E^\nu_{2l}=E^{\nu=0}_{2l}=2l,
\]
whereas the eigenvalues corresponding to the
odd eigenstates $\vert 2l+1\rangle$
are subject to the linear shift,
\[
E^\nu_{2l+1}=E^{\nu=0}_{2l+1}+\nu=2l+1+\nu.
\]
There are two special cases
given by $\nu=2k$ and $\nu=2k+1$
with $k=1,2,\ldots$.
For $\nu=2k$ the spectrum of the system
(\ref{norm}) contains $k$ `holes'
in comparison with the non-deformed (bosonic) case
$\nu=0$: the first $k$ odd energy 
levels with $E=2h-1$, $h=1,\ldots, k,$ are absent from it,
whereas the rest of the spectrum is the same as
for the bosonic system (see Fig. 1a).

On the other hand, 
the odd values of the deformation parameter,
$\nu=2k+1$, supply us with 
supersymmetry. As we have seen, at $k=0$
the Hamiltonian (\ref{norm}) describes the
system in the phase of unbroken supersymmetry
with one singlet state corresponding to
$E=0$ and all other states to be paired in
supersymmetric doublets. 
For the parabosons
of order $p=2(k+1)$, $k=1,2,\ldots$, 
one gets the nonstandard supersymmetry.
It is specified by the presence of $k+1$
singlet states $\vert 2l\rangle$, $l=0,...,k$,
with energies 
$E_{2l}=0,2,\ldots,2k$, whereas other states are
paired in supersymmetric doublets 
with $E_{2(l-k)-1}=E_{2l}=2l$,
$l=k+1,\ldots$ (see Fig. 1b).

One notes that the case of 
the nonstandard $N=1$ supersymmetry with several singlet
states may be treated as
a reduced parasupersymmetry.
Indeed, let us consider the parasupersymmetric system
described by the Hamiltonian 
\begin{equation}
H=b^+b^-+J_3+j.
\label{para1}
\end{equation}
Here $b^\pm$ are the bosonic creation-annihilation
operators and $J_3$ is one of the $su(2)$ generators 
taken in spin-$j$ representation, ${\bf J}^2=j(j+1)$,
with $j=(k+1)/2$, $k=1,2,\ldots$.
The reduction of this spin-$j$ parasupersymmetric system
by two additional linear constraints 
\begin{equation}
(J_3-j)\vert \Psi\rangle=0,\quad
(J_3+j)\vert \Psi\rangle=0
\label{para2}
\end{equation}
(or by one quadratic constraint $(J_3^2-j^2)\vert\Psi\rangle=0$)
results in the spectra corresponding to 
the nonlinear supersymmetry with $k+1$ singlet states
(see Fig. 1c).

\begin{figure}[htb]
\setlength{\unitlength}{1mm}
\noindent
\begin{picture}(135,70)
\put(10,10){\vector(0,1){60}}
\put(2,65){\makebox(10,10)[]{$E$}}
\put(10,10){\line(1,0){10}}
\put(10,10){\circle*{1.5}}
\put(10,20){\circle*{1.5}}
\put(10,30){\circle*{1.5}}
\put(11,35){\circle*{1.5}}
\put(10,40){\circle*{1.5}}
\put(11,45){\circle*{1.5}}
\put(10,50){\circle*{1.5}}
\put(11,55){\circle*{1.5}}
\put(10,60){\circle*{1.5}}
\put(15,0){\makebox(2,2)[]{${\bf a)}$}}

\put(60,10){\vector(0,1){60}}
\put(52,65){\makebox(10,10)[]{$E$}}
\put(60,10){\line(1,0){10}}
\put(60,10){\circle*{1.5}}
\put(60,20){\circle*{1.5}}
\put(60,30){\circle*{1.5}}
\put(60,40){\circle*{1.5}}
\put(62,40){\circle*{1.5}}
\put(60,50){\circle*{1.5}}
\put(62,50){\circle*{1.5}}
\put(60,60){\circle*{1.5}}
\put(62,60){\circle*{1.5}}
\put(65,0){\makebox(2,2)[]{${\bf b)}$}}

\put(110,10){\vector(0,1){60}}
\put(102,65){\makebox(10,10)[]{$E$}}
\put(133,2){\makebox(10,10)[]{$j_3$}}
\put(110,10){\vector(1,0){25}}
\put(110,10){\circle*{1.5}}
\put(110,20){\circle*{1.5}}
\put(115,20){\circle{1.5}}
\put(110,30){\circle*{1.5}}
\put(115,30){\circle{1.5}}
\put(120,30){\circle{1.5}}
\put(110,40){\circle*{1.5}}
\put(115,40){\circle{1.5}}
\put(120,40){\circle{1.5}}
\put(125,40){\circle*{1.5}}
\put(110,50){\circle*{1.5}}
\put(115,50){\circle{1.5}}
\put(120,50){\circle{1.5}}
\put(125,50){\circle*{1.5}}
\put(110,60){\circle*{1.5}}
\put(115,60){\circle{1.5}}
\put(120,60){\circle{1.5}}
\put(125,60){\circle*{1.5}}
\put(120,0){\makebox(2,2)[]{${\bf c)}$}}

\end{picture}
\caption{{\bf a)} Spectrum of the system (\ref{norm}) at $\nu=4$;
{\bf b)} Spectrum of the system (\ref{norm}) at $\nu=5$;
{\bf c)} Spectrum of the reduced parasupersymmetric system (\ref{para1}),
(\ref{para2}) for $j=3/2$  with  the states $j_3=\pm 1/2$ eliminated 
by constraints (\ref{para2}).}
\end{figure}
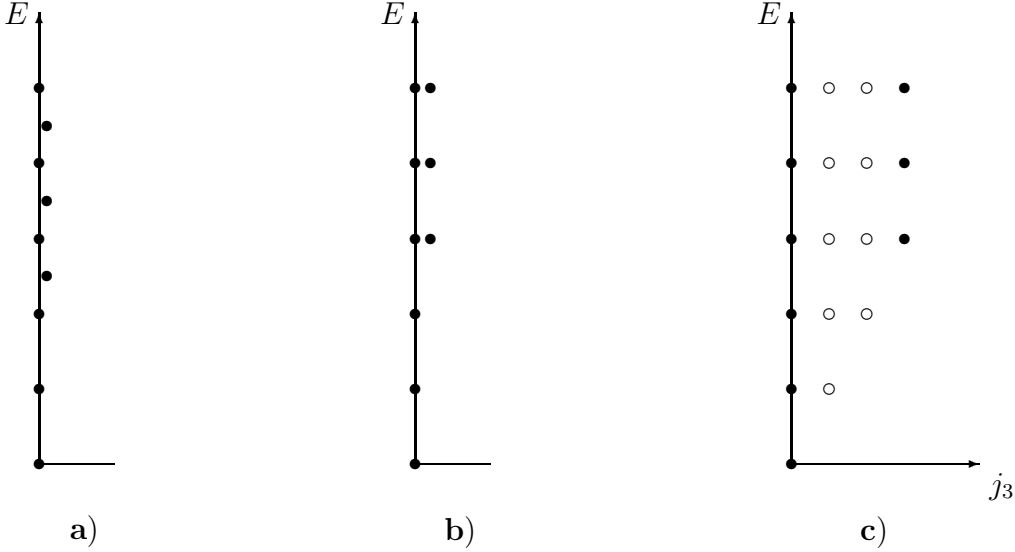

Let us investigate in more detail the 
nature of the observed nonstandard supersymmetry.
First, one constructs the odd
generators 
\begin{equation}
Q^+=a^+\Pi_-,\quad
Q^-=a^-\Pi_+,
\label{q1}
\end{equation}
with $\Pi_\pm=\frac{1}{2}(1\pm R~)$.
Together with the Hamiltonian and grading 
operator $R$ they satisfy the following 
relations:
\[
Q^{\pm 2}=0,\quad
\{R,Q^\pm\}=0,\quad [R,H_n]=0,
\]
\[
\{Q^+,Q^-\}=H_n+k(R-1),\quad
[H_n,Q^\pm]=\mp 2k Q^\pm,\quad
[R,Q^\pm]=\pm 2Q^\pm,\quad R^2=1.
\]
This can be understood as a central extension of
the supersymmetric algebra (\ref{susyal}) with the
grading operator $R$ nontrivially included
in the set of even generators.
Only for $k=0$ the odd generators $Q^\pm$
commute with $H_n$ and are the odd integrals of motion
(supercharges).
In this case they coincide with the operators
$Q^\pm=\frac{1}{2}(Q_1\mp iQ_2)$, where $Q_{1,2}$
are constructed according to (\ref{qq12}) with $W=x-\frac{1}{2x}$.
Therefore, we arrive at the natural question:
what are the supersymmetric generators being
odd integrals of motion which
mutually transform the supersymmetric 
doublet states and what kind of superalgebra do they form?
Such mutually conjugate odd operators
satisfying the relations 
$\{R,\tilde{Q}{}^\pm\}=0$
are 
\begin{equation}
\tilde{Q}{}^+=(a^+)^{2k+1}\Pi_-,\quad
\tilde{Q}{}^-=(a^-)^{2k+1}\Pi_+.
\label{tq1}
\end{equation}
Together with the Hamiltonian $H_n$ they form the following
nonlinear superalgebra:
\begin{equation}
\tilde{Q}{}^{\pm 2}=0,\quad
\{\tilde{Q}{}^+,\tilde{Q}{}^-\}={\cal H}_k(H_n),\quad
[H_n,\tilde{Q}{}^\pm]=0,
\label{n1}
\end{equation}
where 
\[
{\cal H}_k(H_n)=(H_n-2k)(H_n-2k+2)...H_n...(H_n+2k-2)(H_n+2k).
\]
This algebra follows immediately from
the relations 
\[
\tilde{Q}{}^+ \tilde{Q}{}^-={\cal H}_k(H_n)\Pi_+,\quad
\tilde{Q}{}^- \tilde{Q}{}^+={\cal H}_k(H_n)\Pi_-.
\]
The singlet states $\vert 0\rangle_i
\equiv
\vert 2i\rangle$, $i=0,1,\ldots, 2k$,
are annihilated by the odd generators,
$\tilde{Q}{}^\pm\vert 0\rangle_i=0$,
but only the first state $\vert 0\rangle_0=\vert 0\rangle$
is the zero energy ground state: $H_n\vert 0\rangle=0$.
Thus, in the case $\nu=p-1=2k+1$ the supersymmetry
of the single-mode parabosonic system (\ref{norm})
is characterized by the presence of $k+1$ supersymmetric
singlet states and for parabosons of order $p=4,6,\ldots$,
unlike the case of $p=2$,
the corresponding superalgebra is nonlinear.

Realizing the creation-annihilation parabosonic 
operators via Eqs. (\ref{yd}), (\ref{aadef}),
we represent the Hamiltonians $H_n=a^+a^-$ and
supercharges (\ref{tq1}) in the form
\begin{eqnarray}
H_n&=&\frac{1}{2}
\left(
-\frac{d^2}{dx^2}+x^2+\frac{\nu^2}{4x^2}-1+\nu\left(
\frac{1}{2x^2}-1\right)R\right),
\label{c1}\\
\tilde{Q}{}^+&=&(\tilde{Q}{}^-)^{\dagger}=
\frac{1}{2^{3(k+\frac{1}{2})}}
\left(\left(-\frac{d}{dx}+x+\frac{\nu}{2x}\right)(1-R)\right)^{2k+1}
\label{c1q}
\end{eqnarray}
with $\nu=2k+1$.
The system given by the Hamiltonian (\ref{c1})
can be treated as a 2-particle Calogero-like
model with exchange interaction,
where $x$ has a sense of a  
relative coordinate and $R$ has to be understood as an exchange operator. 
The systems of such type were considered by Polychronakos
\cite{poly02,poly,poly03}. 
Therefore, at odd values of the parameter $\nu$
the class of Calogero-like systems 
(\ref{c1}) possesses a hidden supersymmetry which
at $\nu=1$ is the $N=1$ linear supersymmetry
in the unbroken phase, whereas at $\nu=2k+1$, $k=1,2,...$,
the supersymmetry is characterized by the
supercharges being differential operators of order
$2k+1$ satisfying nonlinear superalgebra (\ref{n1}).

\section{The systems with Hamiltonian $H_a=a^-a^+$}

Let us choose now the superpotential in Eqs. (\ref{qq12}), 
(\ref{hams})
in the form 
\[
W=-x-\frac{1}{2x}.
\]
In this case the Hamiltonian can be represented
equivalently as the antinormal product
\begin{equation}
H_a=a^-a^+
\label{antin}
\end{equation}
of the creation-annihilation operators 
of RDHA corresponding 
to the case of parabosons of order $2$ ($\nu=1$).
This antinormal form may be obtained also
from the supersymmetric Hamiltonian (\ref{hamdef})
by putting in it $\epsilon=-$, $\nu=1$.
The corresponding mutually conjugate operators
$Q^\pm=\frac{1}{2}(Q_1\mp i Q_2)$ 
take now the form
\begin{equation}
Q^+=a^+\Pi_+,\quad
Q^-=a^-\Pi_-
\label{qbr}
\end{equation}
instead of supercharges (\ref{q1}) of the system
(\ref{norm}) at $\nu=1$.
Representing the  Hamiltonian (\ref{antin})
in the equivalent form $H_a=N+1+\Pi_+$ in terms
of the number operator and projector $\Pi_+$,
we conclude that it describes 
the system in the phase of  spontaneously broken supersymmetry
characterized by the spectrum 
$E_l=2[l/2]+2$, $l=0,1,\ldots,$ where $[.]$ is the integer part.

Let us follow the same line
as in the previous section and treat the creation-annihilation
operators in the Hamiltonian
(\ref{antin}) as the generators of RDHA 
for the case of generalized parabosons ($\nu>-1$).
Then, taking into account relation (\ref{aan}),
one can represent the Hamiltonian as
\[
H_a=N+1+\nu\Pi_+.
\]
Therefore, the energy levels corresponding to the
odd eigenstates $\vert 2l+1\rangle$ now are stable,
whereas the shift of the levels corresponding to the even 
eigenstates $\vert 2l\rangle$
depends on $\nu$.
Special cases correspond, again,
to the integer values of the deformation parameter,
when RDHA specifies parabosons.
For the parabosons of odd order
$p=\nu+1=2k+1$, $k=1,2,\ldots$,
the spectrum is like in the non-deformed case
$\nu=0$ but with the only difference:
it contains $k$ `holes' in its lower part
in comparison with the bosonic system. 
For odd values $\nu=2k+1$, $k=1,\ldots$,
the spectrum looks like the  
spectrum of usual supersymmetry 
($k=1$) shifted for $\Delta E_l=E_0=2$, or like that
of the nonstandard supersymmetry
($k=2,3,\ldots$).

Let us consider the case of parabosons of even order
$p=\nu +1=4,6,\ldots$,
and shift  the Hamiltonian 
for the constant corresponding to the lowest 
energy level, $H_a\rightarrow H_a^0$,
\begin{equation}
H_a^0=a^-a^+-2.
\label{antin0}
\end{equation}
Then, in the case $p=4$ the spectrum of the system
looks like the spectrum of the superoscillator
\cite{nic} being 
the system in the phase
of unbroken supersymmetry with the lowest singlet
zero energy level and all other levels to be 
supersymmetric doublets.
But what are the corresponding conserved (commuting with
$H_a^0$) odd supergenerators and what superalgebra do they form?
Such mutually conjugate generators for the general case
$\nu=p-1=2k+1$ are given by
\begin{equation}
\tilde{Q}{}^+=(a^{+})^{2k+1}\Pi_+,\quad
\tilde{Q}{}^-=
(a^{-})^{2k+1}\Pi_-.
\label{tq2}
\end{equation}
They satisfy the following relations:
\[
\tilde{Q}{}^+\tilde{Q}{}^-={\cal H}_k(H_a^0)\Pi_-,\quad
\tilde{Q}{}^-\tilde{Q}{}^+={\cal H}_k(H_a^0)\Pi_+,
\]
where
\[
{\cal H}_k(H_a^0)=(H_a^0-2k+2)\cdot (H_a^0-2k+4)\ldots H_a^0
\ldots (H_a^0+2k)\cdot(H_a^0+2k+2).
\]
Using them, we arrive at the following
nonlinear superalgebra which characterizes the system
(\ref{antin0}) at $\nu=2k+1$, $k=1,2,...,$
(or the system given by the Hamiltonian
(\ref{antin}) with the substitution $H^0_a=H_a-2$):
\begin{equation}
\tilde{Q}{}^{\pm 2}=0,\quad
\{\tilde{Q}{}^+,\tilde{Q}{}^-\}={\cal H}_k(H_a^0),\quad
[H_a^0,\tilde{Q}{}^\pm]=0.
\label{qth}
\end{equation}
On the other hand, with the  `first order' odd generators
given by Eq. (\ref{qbr})
the system (\ref{antin0}) can be characterized by the 
following central extension of the $N=1$ superalgebra
with the generators $Q^\pm$, $H_a^0$ and $R$:
\[
Q^{\pm 2}=0,\quad
\{R,Q^\pm\}=0,\quad 
[R,H_a^0]=0,
\]
\[
\{Q^+,Q^-\}=H_a^0-kR+k+2,\quad
[H_a^0,Q^\pm]=\pm 2k Q^\pm,\quad
[R,Q^\pm]=\pm 2Q^\pm,\quad
R^2=1.
\]
Therefore,
in the case $p=4$ the system (\ref{antin0}) 
has the usual spectrum of the system in the phase
of exact supersymmetry, but its odd conserving supergenerators
form with the Hamiltonian the nonlinear superalgebra:
\begin{equation}
\{\tilde{Q}{}^+,\tilde{Q}{}^-\}=
H_a^0(H_a^0+2)(H_a^0+4),\quad
[H_a^0,\tilde{Q}{}^\pm]=0,\quad
\tilde{Q}{}^{\pm 2}=0.
\label{n3}
\end{equation}
Note that here the ground (supersymmetric vacuum) state
is $\vert vac\rangle=\vert 1\rangle$, 
and, so, it is odd, $R\vert 1\rangle=-\vert 1\rangle$,
whereas $\vert 0\rangle$
belongs to the first excited supersymmetric
doublet. Therefore, this system is not included 
in the minimally bosonized version of supersymmetric
quantum mechanics where 
the supersymmetric ground state is always
even \cite{gpz} and this explains the appearance of the
nonlinear superalgebra.
The systems with the Hamiltonian (\ref{antin0}) 
corresponding to $p=6,8,\ldots$
are  characterized by the same property: their
ground state is odd: $\vert vac\rangle=\vert 1\rangle$,
$H^0_a\vert vac\rangle=0$.

The coordinate realization (\ref{yd}), (\ref{aadef}),
results in the following form for the Hamiltonian
$H_a^0=a^-a^+-2$ and supercharges (\ref{tq2}):
\begin{eqnarray}
H_a^0&=&
\frac{1}{2}\left(
-\frac{d^2}{dx^2}+x^2+\frac{\nu^2}{4x^2}-3+\nu\left(
\frac{1}{2x^2}+1\right)R\right),
\label{c2}\\
\tilde{Q}{}^+&=&(\tilde{Q}{}^-)^{\dagger}=
\frac{1}{2^{3(k+\frac{1}{2})}}
\left(\left(-\frac{d}{dx}+x-\frac{\nu}{2x}\right)(1+R)\right)^{2k+1}
\label{c2q}
\end{eqnarray}
with $\nu=2k+1$.
Therefore, among the Calogero-like 2-particle systems 
with exchange interaction given by the Hamiltonian (\ref{c2}) 
the cases corresponding to the odd values of the parameter
$\nu=3,5,7,...$ are characterized by the presence of the
hidden nonlinear supersymmetry,
whereas the system given by the Hamiltonian $H_a^0+2$ at
$\nu=1$ possesses linear supersymmetry (\ref{susyal})
in the spontaneously broken phase.
In particular, as we observed, the 
Calogero-like system (\ref{c2}) corresponding
to $\nu=3$ is a rather peculiar one: its spectrum
coincides exactly with the equidistant 
spectrum of usual linear superoscillator, but the
associated supercharges (\ref{c2q}) satisfy in this case
the nonlinear superalgebra (\ref{n3}).

\section{Realization of parabosons by two identical fermions}

Suppose that we have the system of two identical fermions such
that its space motion is confined to the line whereas their spin
degrees of freedom are not restricted and are given by the
total vector spin operator with components
$J_i=\frac{1}{2}(\sigma_i\otimes 1+1\otimes\sigma_i)$.  Then,
taking into account the Pauli principle and
omitting the dependence on the center of mass coordinate,
$X=\frac{1}{2}(x_1+x_2)$, we describe the two-fermion system by the wave
functions of the form \cite{gpz}
\begin{equation}
\Psi(x)=\chi_s^{j_3}\psi_-^{j_3}(x)+
\chi_a\psi_+(x).
\label{psi}
\end{equation}
Here $j_3=-1,0,1$,
$x=x_1-x_2$ is the relative  coordinate;
$\chi^{+1}_s=\vert +\rangle\vert+\rangle,$
$\chi^{-1}_s=\vert -\rangle\vert-\rangle$
and
$\chi^0_s=\frac{1}{\sqrt{2}}
(\vert +\rangle\vert -\rangle+\vert -\rangle\vert+\rangle)$
are symmetric spin states forming a vector 
triplet, $J_iJ_i\chi^{j_3}_s=2\chi^{j_3}_s$,
$J_3\chi^{j_3}_s=j_3\chi^{j_3}_s$,
whereas 
$\chi_a=\frac{1}{\sqrt{2}}
(\vert +\rangle\vert -\rangle-\vert -\rangle\vert+\rangle)$
is antisymmetric spin-$0$ singlet state,
$J_i\chi_a=0$;
$\psi^{j_3}_-$ are odd functions,
$\psi^{j_3}_-(-x)=-\psi^{j_3}_-(x)$,
and 
$\psi_+$ is an even function,
$\psi_+(-x)=\psi_+(x)$.
The physical operators are those transforming the states of the form
(\ref{psi})
into the states of the same form.
If we assume, moreover,  that there is a
rotational $J_3^2$-symmetry in the spin space,
then the algebra of physical operators
may be generated by the operators \cite{gpz}
\[
{\cal A}_+=f_+\left(x,\frac{d}{dx}\right){\cal O}_+,\quad
{\cal A}_-=f_-\left(x,\frac{d}{dx}\right){\cal O}_-,
\]
where
$f_\pm(-x,-\frac{d}{dx})=\pm
f_\pm(x,\frac{d}{dx})$;
${\cal O}_+=1,\Sigma_3,\Xi_i$, $i=1,2,3$;
${\cal O}_-=\Sigma_1,\Sigma_2$;
\[
\Sigma_1=\frac{1}{2}(\sigma_3\otimes 1-
1\otimes\sigma_3),\quad
\Sigma_2=\frac{1}{2}(\sigma_1\otimes\sigma_2-\sigma_2\otimes\sigma_1),
\quad
\Sigma_3=\frac{1}{2}(\sigma_1\otimes\sigma_1+
\sigma_2\otimes\sigma_2),
\]
\[
\Xi_1=\frac{1}{2}
(\sigma_1\otimes \sigma_1-\sigma_2\otimes \sigma_2),\quad
\Xi_2=\frac{1}{2}
(\sigma_1\otimes\sigma_2
+\sigma_2\otimes\sigma_1),
\quad
\Xi_3=J_3=\frac{1}{2}(1\otimes\sigma_3+\sigma_3\otimes 1).
\]
Operators $\Sigma_i$ and $\Xi_i$ 
satisfy the relations 
$\Sigma_i\Sigma_j={\cal I}\delta_{ij}
+i\epsilon_{ijk}\Sigma_k$,
${\cal I}=1-J_3^2$,
$\Xi_i\Xi_j=J_3^2\delta_{ij}+i\epsilon_{ijk}
\Xi_k,$
$\Sigma_i\Xi_j=0$.
They act on the spin states in the following way:
$\Sigma_1\chi_{s(a)}=
\chi_{a(s)},$
$\Sigma_3\chi_{s(a)}=+(-)\chi_{s(a)},$
$\Sigma_i\chi^\pm=0,$
$\Xi_1\chi^\pm=\chi^\mp,$
$\Xi_3\chi^\pm=\pm\chi^\pm$,
$\Xi_i\chi_{s(a)}=0$,
where we have used the notation
$\chi^\pm=\chi^{\pm 1}_s$,
$\chi_s=\chi^0_s$.

Let us return to the $R$-deformed Heisenberg algebra.
In the coordinate representation an arbitrary
function $\Psi(x)$ may be decomposed into even and odd
parts, $\Psi(x)=\Psi_+(x)+\Psi_-(x)$,
$R\Psi_\pm(x)=\pm\Psi_\pm(x)$.
Even operators $f_+(x,\frac{d}{dx})$
map the even and odd functions into 
the functions of the same parity nature,
whereas odd operators $f_-(x,\frac{d}{dx})$
change the parity of functions.
Then, taking into account the properties of operators
$\Sigma_i$ and $\Xi_i$ specified above,
one can arrive at the following identification 
of the states and operators 
of RDHA on the one hand, and, on the other hand,
of the states and 
nontrivial physical operators in the $j_3=0$
sector of the system of two identical fermions:

\vspace{0.5cm}

\begin{tabular}{|l|c|c|} 
\hline
        & Fermion system ($j_3=0$)  &  RDHA \\ 
  \hline
States    & $\chi_s\psi_-$           &   ${\Psi}_-$  \\
          & $\chi_a\psi_+$           &   ${\Psi}_+$  \\ \hline
Operators & $f_+{\cal I}$            &   $f_+$       \\ 
          & $f_+{\Sigma_3}$          &   $f_+R$      \\
          & $f_-{\Sigma_1}$          &   $f_-$       \\ 
    & $f_-{\Sigma_2}$          &   $if_-R$     \\ \hline
\end{tabular} 

\vspace{0.5cm}

This identification means that the operators 
\begin{equation}
A^\pm=\frac{1}{\sqrt{2}}\Sigma_1\left[
x\mp\left(\frac{d}{dx}-\frac{\nu}{2x}\Sigma_3\right)
\right]
\label{AAF}
\end{equation}
together with the operator $\Sigma_3$ 
are the analogs of the creation-annihilation
operators (\ref{aadef}) and of the reflection operator $R$.
Indeed, the operators $A^\pm$ and $\Sigma_3$
satisfy the relations
\[
[A^-,A^+]=(1+\nu\Sigma_3){\cal I},\quad
\{\Sigma_3,A^\pm\}=0,\quad \Sigma_3^2={\cal I},
\]
which in the sector $j_3=0$
are reduced to the relations of RDHA (\ref{dha}).
Therefore, all the obtained results on the supersymmetries
of single-mode parabosonic systems can be translated to
the $j_3=0$ sector of the system of two identical fermions.
Then
the Hamiltonian 
\begin{equation}
H^f_n=A^+A^-
=
\frac{1}{2}\left[-\frac{d^2}{dx^2}+W_{+,\nu}^2+\nu-1-
(W'_{+,\nu}+\nu-1)\Sigma_3\right]
{\cal I}
\label{hf}
\end{equation}
with $W_{+,\nu}=+x-\frac{\nu}{2x}$
corresponds to Eq. (\ref{norm}).
So,
the system  of two identical fermions
described by the Hamiltonian (\ref{hf})
reveals linear and nonlinear supersymmetries
in $j_3=0$ sector
when the parameter $\nu$ takes the values
$\nu=1$ and $\nu=3,5,7,\ldots$, respectively.
Note that here the states with $j_3=\pm 1$ 
always have zero energy.
Can we modify  the Hamiltonian (\ref{hf})
in such a way that the infinitely degenerated
zero energy levels corresponding to $j_3=\pm 1$ states 
would be unfrozen without
changing the supersymmetric properties
of $j_3=0$ sector?
A natural modification of the Hamiltonian
(\ref{hf}) is 
\begin{equation}
\tilde{H}{}^f_n=
\frac{1}{2}\left[-\frac{d^2}{dx^2}+W_{+,\nu}^2+\nu-1-
\frac{1}{2}(W'_{+,\nu}+\nu-1)\right]-
\frac{1}{4}(W'_{+,\nu}+\nu-1)\cdot\sigma_i\otimes\sigma_i.
\label{th}
\end{equation}
The difference $\Delta=\tilde{H}{}^f_n-H^f_n$
is given by the operator
\[
\Delta=\frac{1}{2}\left[-\frac{d^2}{dx^2}
+W_{+,\nu}^2+\nu-1-(W'_{+,\nu}+\nu-1)\right]\cdot J_3^2,
\]
which acts nontrivially only in the $j_3=\pm 1$ sector.
Moreover, this difference is chosen in such a way
that its action on the states $\chi^{+1}_s\psi^{+1}_-(x)$,
$\chi^{-1}_s\psi^{-1}_-(x)$ coincides
with the action of the operator $H^f_n$
on the state $\chi^0_s\psi_-(x)$.
As a result, the energy levels of Hamiltonian
(\ref{th}) corresponding to all odd eigenstates
(given by odd functions of $x$)
have a multiplicity equal to 3.  
The spectrum 
of the corresponding two-fermion system is
shown in Fig. 2a for the case of $\nu=3$.

The two-fermion Hamiltonian
\begin{equation}
H_a^f=A^-A^+=\frac{1}{2}
\left[-\frac{d^2}{dx^2}+W_{-,\nu}^2-\nu+1-
(W'_{-,\nu}-\nu+1)\Sigma_3\right]{\cal I}
\label{a-a+}
\end{equation}
with $W_{-,\nu}=-x-\frac{\nu}{2x}$
is the analog of the Hamiltonian
(\ref{antin}), and at $\nu=2k+1$ it reveals
supersymmetry in $j_3=0$ sector.
The modified (and shifted for $-2$) Hamiltonian
\begin{equation}
\tilde{H}{}^{0f}_a=
\frac{1}{2}\left[-\frac{d^2}{dx^2}+W_{-,\nu}^2-\nu-3
-\frac{1}{2}(W'_{-,\nu}-\nu+1)\right]-\frac{1}{4}
(W'_{-,\nu}-\nu+1)\cdot\sigma_i\otimes \sigma_i
\label{tha}
\end{equation}
is different from $H^f_a$ in 
\[
\Delta=\frac{1}{2}\left[-\frac{d^2}{dx^2}
+W_{-,\nu}^2-\nu+1-(W'_{-,\nu}-\nu+1)\right]\cdot J_3^2-2.
\]
The energy levels corresponding to odd eigenstates  of the Hamiltonian
(\ref{tha})
(represented by odd functions of $x$)
have also the multiplicity equal to $3$ and 
its spectrum is shown in Fig. 2b
for the case $\nu=5$. 

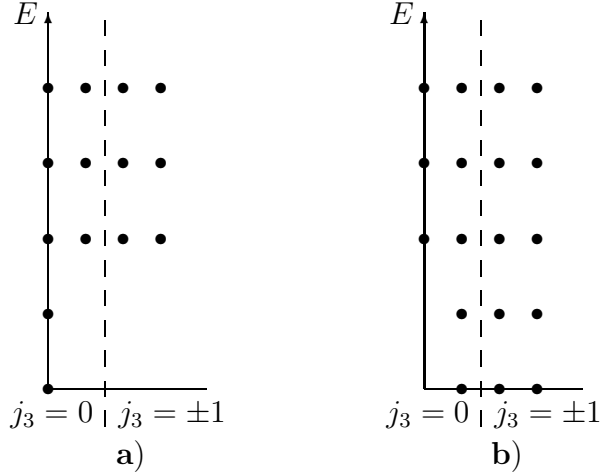
\begin{figure}[htb]
\setlength{\unitlength}{1mm}
\noindent
\begin{picture}(135,60)
\put(50,10){\vector(0,1){50}}
\put(42,55){\makebox(10,10)[]{$E$}}
\put(50,10){\line(1,0){21}}
\put(50,10){\circle*{1.5}}
\put(50,20){\circle*{1.5}}
\put(50,30){\circle*{1.5}}
\put(55,30){\circle*{1.5}}
\put(60,30){\circle*{1.5}}
\put(65,30){\circle*{1.5}}
\put(50,40){\circle*{1.5}}
\put(55,40){\circle*{1.5}}
\put(60,40){\circle*{1.5}}
\put(65,40){\circle*{1.5}}
\put(50,50){\circle*{1.5}}
\put(55,50){\circle*{1.5}}
\put(60,50){\circle*{1.5}}
\put(65,50){\circle*{1.5}}
\put(60,0){\makebox(2,2)[]{${\bf a)}$}}
\put(49,5){\makebox(3,3)[]{${ j_3=0}$}}
\put(65,5){\makebox(3,3)[]{${ j_3=\pm 1}$}}
\multiput(57.5,5)(0,4){14}{\line(0,1){2}}

\put(100,10){\vector(0,1){50}}
\put(92,55){\makebox(10,10)[]{$E$}}
\put(100,10){\line(1,0){21}}
\put(100,30){\circle*{1.5}}
\put(100,40){\circle*{1.5}}
\put(100,50){\circle*{1.5}}
\put(105,10){\circle*{1.5}}
\put(105,20){\circle*{1.5}}
\put(105,30){\circle*{1.5}}
\put(105,40){\circle*{1.5}}
\put(105,50){\circle*{1.5}}
\put(110,10){\circle*{1.5}}
\put(110,20){\circle*{1.5}}
\put(110,30){\circle*{1.5}}
\put(110,40){\circle*{1.5}}
\put(110,50){\circle*{1.5}}
\put(115,10){\circle*{1.5}}
\put(115,20){\circle*{1.5}}
\put(115,30){\circle*{1.5}}
\put(115,40){\circle*{1.5}}
\put(115,50){\circle*{1.5}}
\put(110,0){\makebox(2,2)[]{${\bf b)}$}}
\multiput(107.5,5)(0,4){14}{\line(0,1){2}}
\put(99,5){\makebox(3,3)[]{${ j_3=0}$}}
\put(115,5){\makebox(3,3)[]{${ j_3=\pm 1}$}}

\end{picture}
\caption{{\bf a)} Spectrum of the two-fermion system (\ref{th}) at $\nu=3$;
{\bf b)} Spectrum of the two-fermion system (\ref{tha}) at $\nu=5$.}
\end{figure}

Thus, the supersymmetries of the single-mode parabosonic systems
may be realized in the sector
$j_3=0$ of the system of two identical fermions.
The modified Hamiltonians (\ref{th}) and (\ref{tha})
have the form of the Hamiltonians
corresponding to the simplest 2-particle case
of spin chain models with inverse-square interaction
\cite{spch1}.
These Hamiltonians 
unfreeze the energy levels of the states
with $j_3=\pm 1$ (which for the Hamiltonians
(\ref{hf}) and (\ref{a-a+}) turn out to be 
infinitely degenerated states of zero energy)
leading to the multiplicity equal to $4$ for higher degenerated states.

\section{Nonlinear supersymmetry in boson-fermion system}

Nonlinear supersymmetry specific to the pure parabosonic 
system may also be realized by the boson-fermion
system. To construct such a boson-fermion
system with nonlinear supersymmetry,
it is convenient to start from the 
(pseudo)classical level using the ideas 
of ref. \cite{clas}.
So, let us consider
a nonrelativistic particle of unit mass
carrying the odd degrees of freedom described
by two Grassmann variables 
$\theta_a$, $a=1,2$.
The general Lagrangian of such a system
is 
\[
L=\frac{1}{2}\dot{x}{}^2-V(x)-L(x)\Gamma +\frac{i}{2}
\theta_a\dot{\theta}_a,
\]
where $V(x)$ and $L(x)$ are two arbitrary functions
and $\Gamma=-i\theta_1\theta_2=\theta^+\theta^-,$
$\theta^\pm=\frac{1}{\sqrt{2}}(\theta_1\pm i\theta_2)$.
The nontrivial Poisson-Dirac brackets for the system are
$\{x,p\}=1$, $\{\theta_a,\theta_b\}=-i\delta_{ab}$,
and the Hamiltonian $H=\frac{1}{2}p^2+V(x)+L(x)\Gamma$
generates the following equations of motion:
\[
\dot{x}=p,\quad
\dot{p}=-V'(x)-L'(x)\Gamma,\quad
\dot{\theta}{}^\pm=\pm iL(x)\theta^\pm.
\]
Let us construct the oscillator-like variables
\[
B^\pm=\frac{1}{\sqrt{2}}(W(x)\mp ip)
\]
with some function $W(x)$.
We find that 
\[
\dot{B}^\pm=\pm iW'B^\pm\pm\frac{i}{\sqrt{2}}
(V'-WW')\pm \frac{i}{\sqrt{2}}L'\Gamma.
\]
As a consequence, the odd quantities
\begin{equation}
Q^\pm_k=(B^\pm)^k\theta^\mp,
\label{qodd}
\end{equation}
where $k$ is some natural number,
satisfy the following equations of motion:
\[
\dot{Q}{}^\pm_k=\pm iQ^\pm_k(kW'-L)
\pm \frac{i}{\sqrt{2}}k(V'-WW')\theta^\mp.
\]
They 
will be odd integrals of motion if we choose
$V'=\frac{1}{2}(W^2)'$, and put $L=kW'$.
Therefore, when the functions $V(x)$ and $L(x)$
are related as $L(x)=kW'(x)$,
$V(x)=\frac{1}{2}W^2(x)+C$,
where $C$ is a real constant, 
then the odd integrals (\ref{qodd})
are the integrals of motion additional to
$H$ and $\Gamma$.
Together with the Hamiltonian 
\begin{equation}
H_k=\frac{1}{2}(p^2+W^2)
+kW'\Gamma+C
\label{cln}
\end{equation}
they form the following nonlinear superalgebra:
\begin{equation}
\{Q^+_k,Q^-_k\}=-i(H_k-C)^k,\quad
\{H_k,Q^\pm_k\}=\{Q^+_k,Q^+_k\}=\{Q^-_k,Q^-_k\}=0,
\label{noncl}
\end{equation}
whereas $\Gamma$ is the classical analog of the grading 
operator, $\{\Gamma,Q^\pm_k\}=\{\Gamma,H_k\}=0$.
So, the class of (pseudo)classical
systems revealing nonlinear supersymmetry algebra
is given by an arbitrary function $W(x)$:
\begin{equation}
L=\frac{1}{2}\left(\dot{x}{}^2-W^2(x)+
ikW'(x)\epsilon_{ab}\theta_a\theta_b +
i\theta_a\dot{\theta}_a\right),
\label{kl}
\end{equation}
where we have put $C=0$.

Let us consider the simplest example
corresponding to the case of
linear superpotential $W=x$.  
The quantum Hamiltonian of such a system
is the sum of the Hamiltonians 
of harmonic bosonic and fermionic
oscillators with the
fermionic frequency being the integer 
multiple of the bosonic one:
\begin{equation}
H_k=H_b+kH_f,\quad
H_b=b^+b^-,\quad
H_f=f^+f^-.
\label{hk}
\end{equation}
In the case $k=1$ the system is a superoscillator
\cite{nic} for which the quantum analog 
of superalgebra (\ref{noncl}) is given by Eq. (\ref{susyal}).
For $k=2,3,\ldots$ the quantum analog
of superalgebra (\ref{noncl}) for the system (\ref{hk})
takes the form similar to the nonlinear superalgebra
of parabosonic system (\ref{n1}) or (\ref{qth}):
\[
\{Q^+_k,Q^-_k\}=H(H-1)\ldots(H-k+1).
\]
The boson-fermion system (\ref{hk}) possessing the nonlinear
supersymmetry of the type we have revealed in the single-mode
parafermionic systems
is analogous to the system of bosonic oscillators
with rational ratio of frequencies,
which, in turn, is characterized by the nonlinear
finite $W$-symmetry \cite{gr,n3}.

For the superpotential $W(x)$ different from
the linear one the commutator of operators $B^+$ and $B^-$ is nontrivial
and in general case this leads to the quantum anomalies
destroying the classical nonlinear supersymmetry \cite{kppre}.

The `hole' structure specific for
the parabosonic systems (\ref{norm}) and (\ref{antin}) at $\nu=2k$
may also be reproduced by the simplest boson-fermion system in the obvious way:
the spectrum of the Hamiltonian $H=2[b^+b^-+(k+\frac{1}{2})f^+f^-]$
reveals $k$ `holes' in comparison with the spectrum of the bosonic
number operator $H_b=b^+b^-.$

\section{Discussion and outlook}

To conclude, let us discuss the obtained results
and indicate the problems that deserve further attention.

We have found that the single-mode systems of  parabosons of even order
with the Hamiltonians $H_n=a^+a^-$ and $H_a=a^-a^+$ possess the usual
or nonstandard supersymmetries.  
These supersymmetries are characterized by the superalgebra 
of the  form (\ref{susyal}) linear in the Hamiltonian,
or by the nonlinear superalgebras (\ref{n1}) or (\ref{qth}).
The peculiar nature of the revealed supersymmetries
is encoded in the structure of the corresponding 
conserved odd operators (\ref{tq1}) and
(\ref{tq2}).
They contain the projectors $\Pi_\pm$, which in terms of
the parabosonic creation-annihilation operators may be represented  as
\[
\Pi_+=\cos^2 F,\quad
\Pi_-=\sin^2F,\quad F=\frac{\pi}{4}\{a^+,a^-\},
\]
where we have 
used Eqs. (\ref{N}) and (\ref{R})
and taken into account that $\nu=2k+1$.
Therefore, the conserved odd supercharges are essentially
nonlinear operators of $a^\pm$.

The $R$-deformed Heisenberg algebra, with the help
of which we have analyzed hidden supersymmetries
of the single-mode parabosonic systems, 
can be characterized in accordance with
Eqs. (\ref{aan}), (\ref{R}) by the relations
\begin{equation}
a^+a^-=F(N),\quad
a^-a^+=F(N+1),
\label{chf}
\end{equation}
where $F(N)\equiv N(-1)^N+\nu\sin^2\frac{\pi N}{2}$.
The discussed supersymmetric structure 
of the parabosonic Hamiltonians $H_n$ and $H_a$ is 
hidden in the relation 
\begin{equation}
F(2n+1)=F(2n+\nu+1),\quad
\nu=2k+1,\quad k=0,1,\ldots.
\label{period}
\end{equation}
RDHA can be treated  as a particular
case of the
so called generalized
deformed oscillator algebra \cite{das}
given by the relations (\ref{chf}), by
commutation relations characterizing
the number operator, $[N,a^\pm]=\pm a^\pm$,
and by the structure function $F(0)=0,$ $F(n)>0$,
$n=1,...$, being an analytic function. 
Therefore, it is interesting to investigate 
in the context of hidden supersymmetries the general 
class of single-mode generalized deformed oscillator 
systems whose characteristic functions
have the property of degeneracy analogous to (\ref{period}).
The first results in this direction have been  obtained recently
by observation of the hidden linear and nonlinear supersymmetries  
in parafermionic type  systems (whose characteristic functions
are characterized by additional relation $F(p+1)=0$) \cite{paraf}.

We saw that the realization of 
the creation-annihilation parabosonic 
operators via Eqs. (\ref{yd}), (\ref{aadef})
results in representation
of the Hamiltonians $H_n=a^+a^-$ and
$H_a^0=a^-a^+-2$ in the form (\ref{c1}), (\ref{c2}),
which may be treated as 2-particle Calogero-like
models with exchange interaction.
The obtained results say that the 
odd values of the ``coupling constant'' 
$\nu$ characterize the special subclass
of such Calogero-like models possessing
the hidden supersymmetries.
These supersymmetries are described generally
by nonlinear superalgebras. To our knowledge,
the nonlinear supersymmetries is a new feature
of Calogero-like systems with exchange interaction
not observed earlier. Therefore,
it is interesting to generalize 
the systems (\ref{c1}), (\ref{c2}) 
for the $n$-particle case, $n>2$, (by introducing the set of
corresponding exchange operators \cite{poly,poly02})
to reveal the possible hidden 
nonlinear supersymmetric structure
in n-particle Calogero-like models with exchange
interaction.
As in the $2$-particle case (\ref{c1}),
(\ref{c2}), the peculiar nature of such supersymmetries
has to be encoded in the `bosonized' structure of the supercharge
operators which have to depend explicitly on
exchange operators and generally will be 
higher order operators in derivatives 
$\partial/\partial x_i$ (see Eqs. (\ref{c1q}), (\ref{c2q})).

We have observed that the spectrum of supersymmetric 
parabosonic systems characterized by the
presence of several supersymmetric singlet states
may be reproduced by the reduced parasupersymmetry.
The interesting problem is finding the classical
analog for the reduced parasupersymmetry.
The  corresponding classical systems
may be realized with the usage of paraGrassmann variables, and the 
reduction conditions (\ref{para2}) will appear 
as the quantum analogs of classical first-class constraints.
Such a scheme of obtaining nonlinear supersymmetries
by reduction of the appropriate parasupersymmetric systems
will be analogous to the Poisson reduction of the Kirillov
Poisson structure, 
with the help of which the nonlinear
finite $W$-algebras are constructed \cite{n3}.

We have shown that the $R$-deformed Heisenberg algebra and the
associated supersymmetries of the single-mode parabosonic systems are
realizable in the $j_3=0$ sector of the system of two identical
fermions. 
The correspondence is one-to-one and the crucial role
is played here by the Pauli principle.  On the other hand, by
analogy with relations (\ref{AAF}), one can realize the
operators $a^\pm$ and $R$ as
\begin{equation}
a^\pm=\frac{1}{\sqrt{2}}\sigma_1\left[
x\mp\left(\frac{d}{dx}-\frac{\nu}{2x}\sigma_3\right)
\right],\quad R=\sigma_3.
\label{aas}
\end{equation}
Operators (\ref{aas}) satisfy the relations of RDHA (\ref{dha}),
but such a realization is reducible.  This is clear just from
the observation that the operator ${\cal R}=(-1)^N\sigma_3$,
with $N$ constructed according to Eq. (\ref{N}),
commutes with all the operators (\ref{aas}).

Though we have seen that the nonlinear superalgebra
may be reproduced via the appropriate modification
of the classical analog of the usual supersymmetric 
quantum mechanics, nevertheless the quantum
systems obtained after quantization are different
from the supersymmetric pure parabosonic systems.
Indeed, in supersymmetric
quantum mechanical systems there are 
independent fermionic $f^\pm$ and bosonic 
$b^\pm$ operators. 
Proceeding from the parabosonic operators 
$a^\pm$ one can also construct the operators
$\tilde{b}{}^\pm$ and $\tilde{f}{}^\pm$
satisfying the (anti)commutation relations
$[\tilde{b}{}^-,\tilde{b}{}^+]=1$,
$\{\tilde{f}{}^-,\tilde{f}{}^+\}=1$,
$\tilde{f}{}^{\pm2}=0$,
but they will not be independent:
$[\tilde{b}{}^\pm,\tilde{f}{}^\pm]\neq 0$
(see refs. \cite{mp02,mp01}).
One can show \cite{kppre}
that the parabosonic systems with nonlinear supersymmetry
may be related to the nonlinear supersymmetric 
quantum mechanical analog of  
classical systems (\ref{cln}) in the same way 
as the minimally bosonized version of supersymmetric 
quantum mechanics with its linear supersymmetry
(\ref{susyal}) is related to the usual supersymmetric
quantum mechanics, i.e. 
via the special nonlocal unitary transformation applied to the latter
with subsequent reduction to a subspace of one of
the eigenstates
of the total reflection operator \cite{gpz}.

Among the quantum systems corresponding to the
modified classical analog of supersymmetric quantum mechanics
(\ref{kl}), one can find anomaly-free subclasses of the systems
with nonlinear superpotential \cite{kppre}. Investigating them, 
we hope to establish a relationship between nonlinear supersymmetry
observed here and the known construction of the hierarchy 
of Hamiltonians related by linear supersymmetry, which plays
important role in some integrable systems \cite{susyqm}. Since
the nonlinear supersymmetry of the boson-fermion system
(\ref{hk}) is a super-analog of the nonlinear
symmetry of the system of bosonic oscillators
with rational ratio of frequencies \cite{gr,n3},
investigation of the anomaly-free quantum 
systems corresponding to the classical systems (\ref{kl})
with nonlinear superpotential $W(x)$ attracts also attention
from the view-point of nonlinear finite $W$-symmetries.

\vskip5mm
{\bf Acknowledgements}
\vskip5mm

I am grateful to Jorge Zanelli for discussions.
The work has been supported in part by 
grant 1980619 from FONDECYT (Chile) and by DICYT (USACH).

\appendix

\section{Appendix: Irreducible representations
of RDHA}

Here we give some information on RDHA and its representations,
which is necessary for understanding the main text.
The details and further references
may be found in \cite{def}.

As a consequence of Eqs. (\ref{dha}),
the operators $a^\pm$ obey the trilinear
commutation relations
\begin{equation}
[\{a^-,a^+\},a^\pm]=\pm 2 a^\pm.
\label{tri}
\end{equation}
If one defines the vacuum state as 
$a^-\vert 0\rangle=0$, $R\vert 0\rangle=\vert 0\rangle$,
$\langle 0\vert 0\rangle=1$,
than Eq. (\ref{dha}) gives the relation
\begin{equation}
a^-a^+\vert 0\rangle=(1+\nu)\vert 0\rangle.
\label{avac}
\end{equation}
For integer values of the deformation parameter
$\nu=p-1$, $p=1,2,\ldots$, the relation (\ref{avac})
together with the trilinear relations
(\ref{tri}) specifies the creation-annihilation
operators of the parabosons of order $p$,
whereas for non-integer $\nu>-1$ (see below)
the operators $a^\pm$ satisfying relations
(\ref{dha}) may be treated as the creation-annihilation
operators of generalized parabosons \cite{param}.

The trilinear relations (\ref{tri})
can be represented equivalently as
$\{G,a^\pm\}=0$,
$G=[a^-,a^+]-1$.
As a consequence, $[G^2,a^\pm]=0$,
and one concludes that in irreducible representation
$G^2=const$. If, moreover, $G$ is hermitian operator,
then $G=\nu R$ with $\{R,a^\pm\}=0$,
$R^2=1$ and $\nu\in {\bf R}$.
In other words, the trilinear relations themselves lead
to RDHA (\ref{dha}).

Relations (\ref{tri}) mean that up to the 
additive constant the number operator
is given by a half of anticommutator of $a^+$ and $a^-$.
The commutator from Eq. (\ref{dha})
together with the specified above definition of vacuum state
fixes the additive constant and we 
get the number operator
\begin{equation}
N=\frac{1}{2}\{a^+,a^-\}-\frac{1}{2}(1+\nu).
\label{N}
\end{equation}
In its terms, the reflection operator can be realized as
\begin{equation}
R=(-1)^N=\cos \pi N.
\label{R}
\end{equation}
With relations (\ref{dha}) and (\ref{N}),
the normal and antinormal products of the creation-annihilation
operators  can be given in terms of the number operator as follows:
\begin{equation}
a^+a^-=N+\nu\Pi_-,\quad
a^-a^+=N+1+\nu\Pi_+,
\label{aan}
\end{equation}
where 
$
\Pi_\pm=\frac{1}{2}(1\pm R)
$
are the projector operators, 
$\Pi_-^2=\Pi_-$, $\Pi_+^2=\Pi_+$,
$\Pi_+\Pi_-=0$, $\Pi_++\Pi_-=1$.

{}From the requirement $\langle n\vert n \rangle>0$,
$\vert n\rangle\propto (a^+)^n\vert 0\rangle$,
$n=0,1,2,\ldots$, 
it follows that RDHA has unitary
Fock-type representations when $\nu>-1$.
In this case the orthonormal 
eigenstates of the number operator
are given by $\vert n\rangle=C_n (a^+)^n\vert 0\rangle$,
$C_n=([n]_\nu!)^{-1/2}$,
$[n]_\nu!=\prod_{l=1}^n[l]_\nu,$
$[l]_\nu=l+\frac{1}{2}(1-(-1)^l)\nu$.
For integer values $\nu=k$, $k=0,1,\ldots$,
these representations  
are equivalent to the Fock representation for parabosons
of the corresponding order $p=k+1$.

Let us turn to the Schr${\rm \ddot{o}}$dinger (coordinate) representation.
Here, in correspondence with Eq.
(\ref{aadef}), we have
$\sqrt{2}a^-=x+\frac{d}{dx}-\frac{\nu}{2x}R$.
Then the ground state, specified by the conditions
$R\Psi_0=\Psi_0$, $a^-\Psi_0=0$,
is given by the function
$\Psi_0(x)=\vert x\vert^{\nu/2}\exp(-x^2/2){\cal N}$,
where ${\cal N}$ is a normalization constant.
With the scalar product defined by 
\begin{equation}
(\Psi_1,\Psi_2)=\int_{-\infty}^{+\infty}\Psi_1^*(x)\Psi_2(x)dx,
\label{scal}
\end{equation}
we get 
\[
(\Psi_0,\Psi_0)=2{\cal N}^2\int_0^\infty
\vert x\vert^\nu\exp(-x^2)dx.
\]
The integral is converging (at lower limit)
iff $\nu>-1$, i.e. the ground state is given
by the normalizable function only for $\nu>-1$.
Therefore, one concludes that
in Fock and Schr${\rm \ddot{o}}$dinger representations  
singularity of $\nu\leq -1$ reveals itself in different ways.
In Fock representation for $\nu=-1$ the state
$\vert 1\rangle$ has a zero norm,
$\langle 1\vert 1\rangle=0$,
whereas for $\nu<-1$ there appear the states
with negative norm (in particular,
$\langle 1\vert 1\rangle<0$).
In Schr${\rm \ddot{o}}$dinger representation,
the ground state is not normalizable for $\nu\leq -1$
with respect to the usual scalar product (\ref{scal}).

For the sake of completeness we note that the algebra
(\ref{dha}) reveals some universality \cite{def,3ds}: 
for the special values 
$\nu=-(2p+1)$, $p=1,2,\ldots$,
it has non-unitary finite-dimensional representations
characterized by the relations $(a^\pm){}^{2p+1}=0$.
One can pass over to the hermitian conjugate operators 
$f^-=a^-$, $f^+=Ra^+$. They 
generate the parafermionic algebra of order $p=2$
(for $\nu=-3$), or supply us with the deformation of
parafermions of order $p=4,6,\ldots$, which contains the internal
$Z_2$-grading structure:
\[
[[f^+,f^-]f^\pm]=2(2I_3\mp 1)R f^\pm,
\]
where
\[
R=(-1)^{I_3+p},\quad
I_3=\frac{1}{2}[f^+,f^-]\cdot (-1)^{\frac{1}{2}[f^+,f^-]+p}.
\]

\end{document}